\documentclass{emulateapj}

\slugcomment{Accepted for publication in the Astrophysical Journal} 

\shorttitle{Cold $r$-Process}

\shortauthors{Wanajo}

\begin{document}

\title{Cold \lowercase{$r$}-Process in Neutrino-Driven Winds}

\author{Shinya Wanajo}

\affil{Department of Astronomy, School of Science,
   University of Tokyo, Bunkyo-ku, Tokyo, 113-8654, Japan;
   wanajo@astron.s.u-tokyo.ac.jp}

\begin{abstract}
The $r$-process in a low temperature environment is explored, in which
the neutron emission by photodisintegration does not play a role (cold
$r$-process). A semi-analytic neutrino-driven wind model is utilized
for this purpose. The temperature in a supersonically expanding
outflow can quickly drop to a few $10^8$~K, where the ($n,
\gamma$)-($\gamma, n$) equilibrium is never achieved during the heavy
$r$-nuclei synthesis. In addition, the neutron capture competes with
the $\beta$-decay owing to the low matter density. Despite such
non-standard physical conditions for the cold $r$-process, a
solar-like $r$-process abundance curve can be reproduced. The cold
$r$-process predicts, however, the low lead production compared to
that expected in the traditional $r$-process conditions, which can be
a possible explanation for the low lead abundances found in a couple
of $r$-process-rich Galactic halo stars.
\end{abstract}


\section{Introduction}
It has been generally believed that the high entropy supernova ejecta
(neutrino-driven wind) is the likely site for the rapid
neutron-capture ($r$-) process \citep[][]{Woos92, Meye92, Woos94,
Taka94}. In the previous nucleosynthesis calculations related to the
neutrino-driven wind scenario, the $r$-process has been examined in
sufficiently high temperature conditions, such that $T_9 \gtrsim 1.0$
(where $T_9$ is the temperature in units of $10^9$~K). This physical
condition, as well as the high neutron number density ($N_n >
10^{20}$~cm$^{-3}$), has long been considered as the physical
requirements to account for the solar $r$-process abundance curve
\citep{Math90, Krat93}. However, if the neutrino-driven outflow
becomes supersonic, the temperature can quickly fall below $T_9 = 1$
\citep{Arco07}.

\citet{Wana02} have examined the $r$-process in such supersonically
expanding outflows by introducing the ``freezeout temperature''
$T_\mathrm{f}$ (or $T_\mathrm{9f}$ in units of $10^9$~K) that mimics
the abrupt wind deceleration by the preceding supernova ejecta. In
their calculations, the solar $r$-process curve around the third
abundance peak ($A = 195$) was best reproduced with
$T_\mathrm{9f}\approx 1.0$, and the lower $T_\mathrm{9f}$ (down to
0.6) resulted in forming a lower shifted peak. As shown in
\citet{Arco07}, however, it is quite possible that the wind
temperature drops to a few $10^8$~K before deceleration.

In this \textit{Letter}, the $r$-process in such a low temperature
environment (hereafter ``cold $r$-process'') is explored. A
semi-analytic neutrino-driven wind model \citep{Dunc86, Qian96,
Card97, Otsu00, Wana01, Thom01} is used to obtain the thermodynamic
histories of outflows (\S~2). Nucleosynthesis calculations in the
winds are performed with various $T_\mathrm{9f}$ including a rather
low value that has not been considered in previous works
(\S~3). Fundamental behaviors of the cold $r$-processing are then
discussed in some detail. In \S~4, a possible abundance feature in the
cold $r$-process is discussed. Finally, implications of this study are
presented (\S~5).

\begin{figure}
\epsscale{1.0}
\plotone{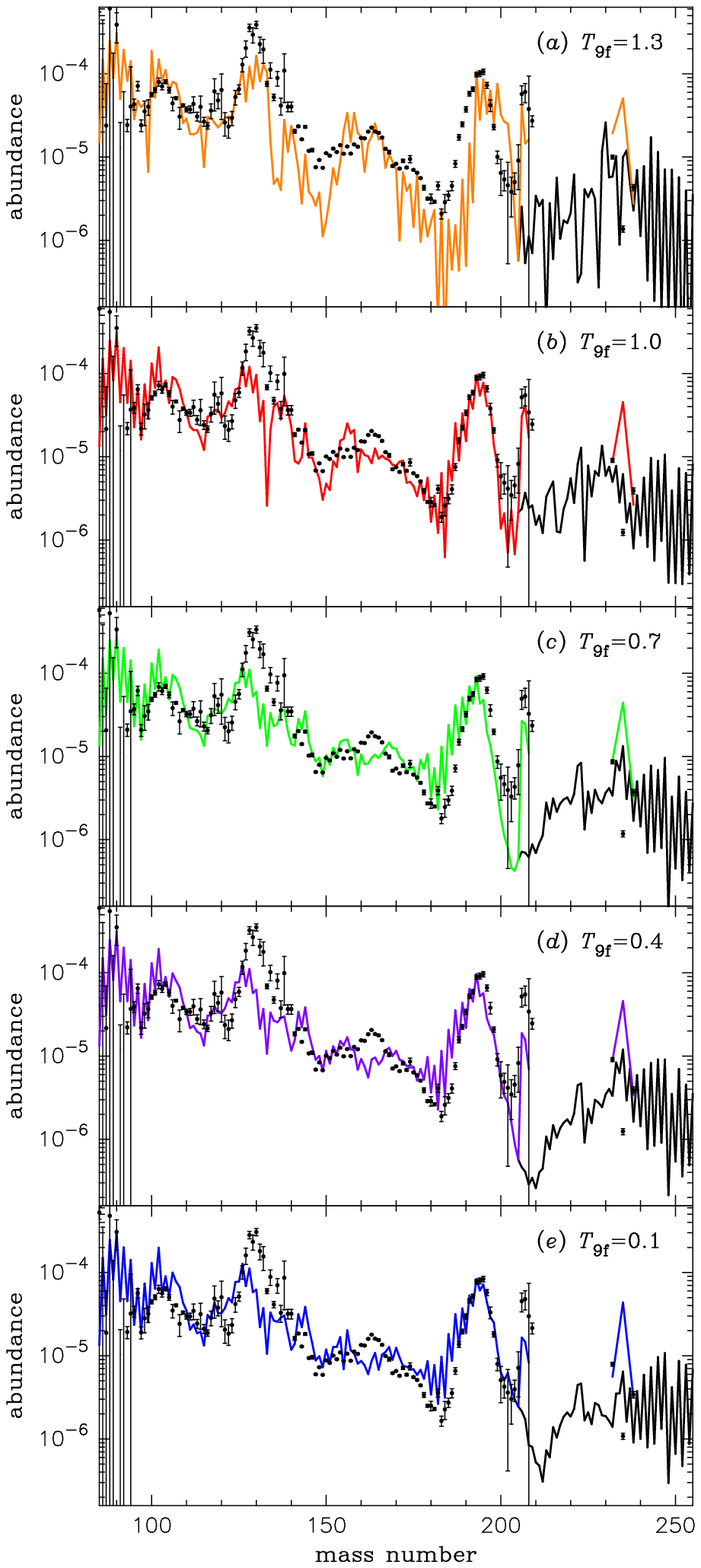}
\caption{Comparison of the mass-averaged yields (\textit{line}) for
$T_\mathrm{9f} = $ (a) 1.3, (b) 1.0, (c) 0.7, (d) 0.4, and (e) 0.1
with the solar $r$-process abundances \citep[][\textit{dots}]{Kapp89}
scaled at the height of the third $r$-process peak ($A \approx 195$),
as functions of mass number. Color and black lines are the yields
before and after $\alpha$-decay, respectively.}
\end{figure}

\begin{figure}
\epsscale{1.0}
\plotone{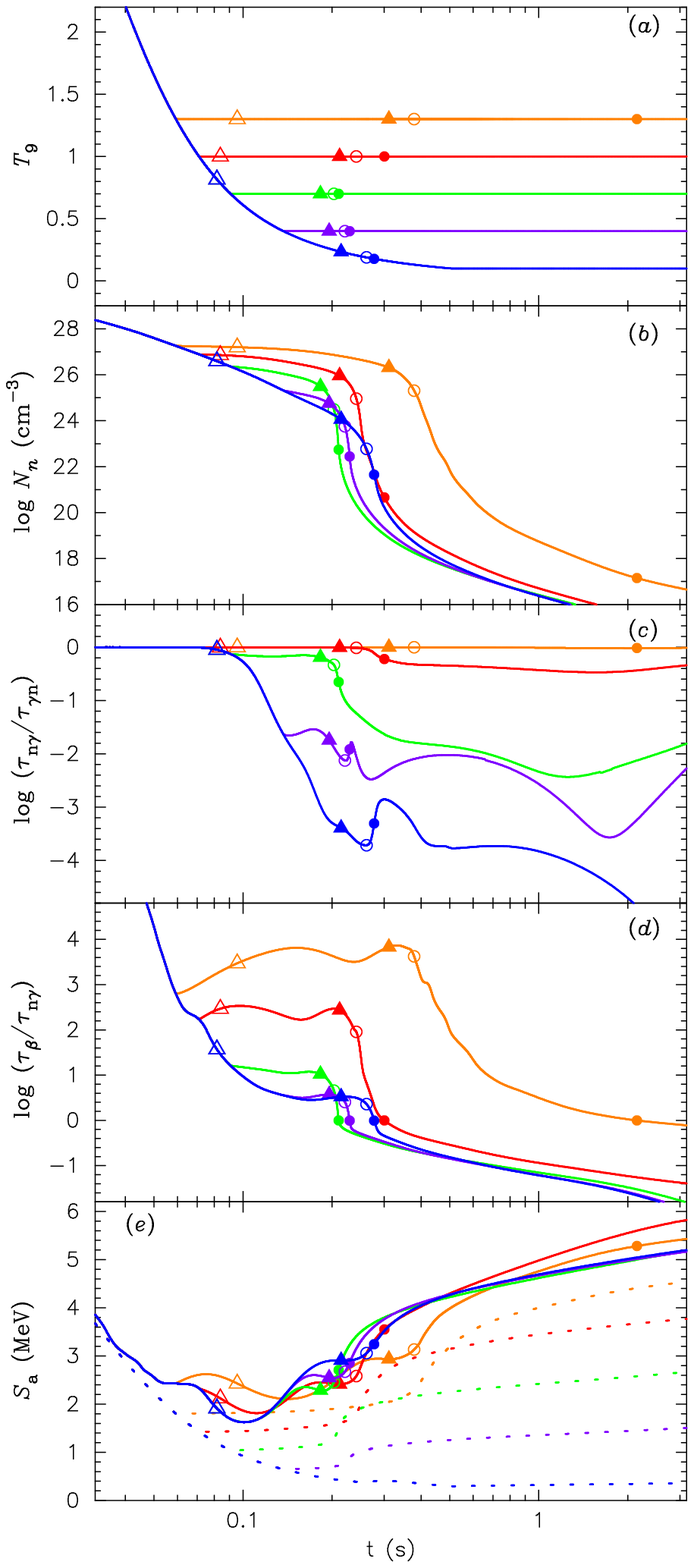}
\caption{Time variations of (a) $T_9$, (b) $N_n$, (c)
$\tau_{n\gamma}/\tau_{\gamma n}$, (d) $\tau_\beta/\tau_{n\gamma}$, and
(e) $S_\mathrm{a}$ for $L_\nu = 2 \times 10^{51}\, \mathrm{erg\
s}^{-1}$ as functions of time. The assigned color in each
$T_\mathrm{9f}$ case is as the same as in Figure~1. The times
corresponding to the second ($A = 130$) and third ($A = 195$) peak
formation, $n$-exhaustion, and freezeout are marked by \textit{open
and filled triangles}, and \textit{open and filled circles},
respectively. The \textit{dotted lines} in the bottom panel show
$S_\mathrm{a}^0$ (see text).}
\end{figure}

\begin{figure}
\epsscale{1.0}
\plotone{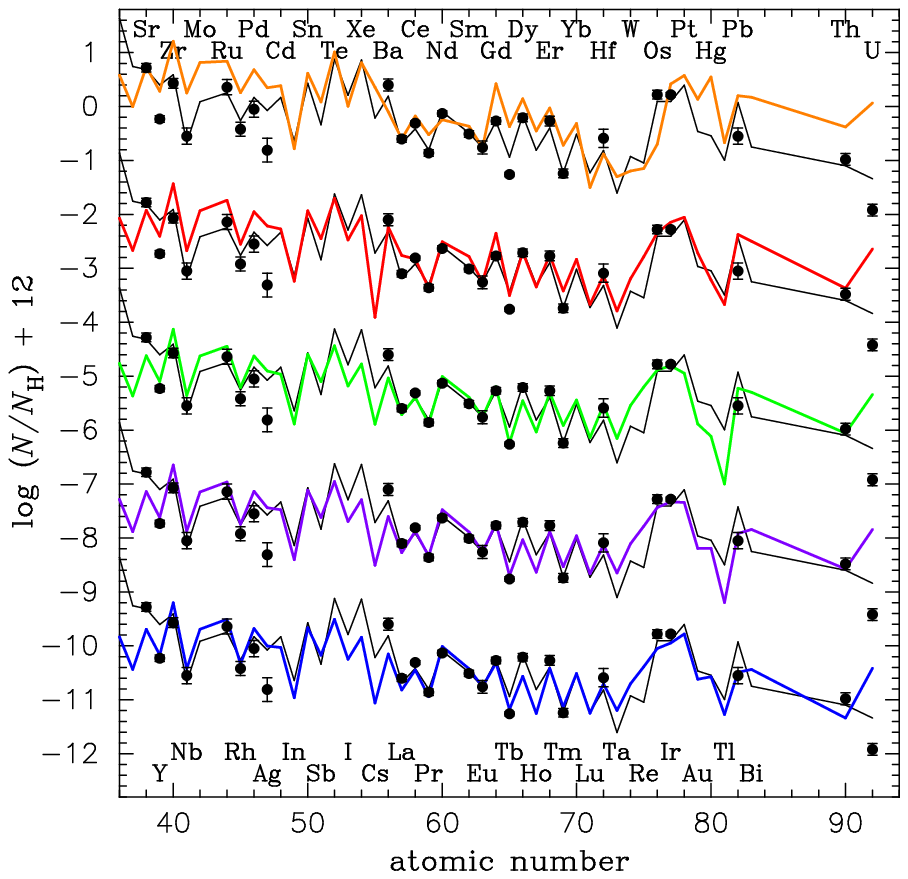}
\caption{Comparison of the mass-averaged yields (\textit{colored
lines}) with abundances for CS~31082-001
\citep[\textit{circles},][]{Hill02, Ivar03, Plez04} and the solar
$r$-process curve \citep[\textit{black lines},][]{Burr00} scaled at
the Eu ($Z = 63$) value, as functions of atomic number. The assigned
color in each $T_\mathrm{9f}$ case is as the same as in Figure~1. The
vertical scale for the uppermost set is true, and the others are
scaled downward for display purposes.}
\end{figure}

\section{Nucleosynthesis in Winds}
The wind trajectories are obtained using the semi-analytic,
spherically symmetric, general relativistic model of neutrino-driven
winds \citep{Wana01, Wana02}. The rms average neutrino energies are
taken to be 10, 20, and 30~MeV, for electron, anti-electron, and the
other flavors of neutrinos, respectively. The mass ejection rate at
the neutrino sphere $\dot M$ is determined so that the wind becomes
supersonic through the sonic point.

The neutron star mass is taken to be $1.4\, M_\odot$. The radius of
neutrino sphere is assumed to be $R_\nu (L_\nu) = (R_{\nu 0} - R_{\nu
1}) (L_\nu/L_{\nu 0}) + R_{\nu 1}$ as a function of neutrino
luminosity $L_\nu$, where $R_{\nu 0} = 15\, \mathrm{km}$, $R_{\nu 1} =
10\, \mathrm{km}$, and $L_{\nu 0} = 1 \times 10^{52}\, \mathrm{ergs\
s}^{-1}$. This mimics the early time evolution of $R_\nu$ in
hydrodynamic simulations \citep{Woos94}. The wind trajectories are
calculated for 39 constant $L_\nu = (10-0.5) \times 10^{51}\,
\mathrm{erg\ s}^{-1}$ with the interval of $0.25 \times 10^{51}\,
\mathrm{erg\ s}^{-1}$. In order to link the series of constant $L_\nu$
trajectories to realistic time-evolving winds, $L_\nu$ is defined as
$L_\nu (t_\mathrm{pb}) = L_{\nu 0} (t_\mathrm{pb}/t_0)^{-1}$, where
$t_\mathrm{pb}$ is the post bounce time and $t_0 = 1.0$~s.

The temperature and density in each wind are set to be constant when
$T_9$ decreases to a given freezeout temperature $T_\mathrm{9f}$
(Fig.~2a), in order to mimic the wind deceleration by the preceding
ejecta. In this study, $T_\mathrm{9f}$ is taken to be 1.3, 1.0, 0.7,
0.4, and 0.1.

The nucleosynthetic yields in each wind are obtained by solving an
extensive nuclear reaction network code. The network consists of 6300
species between the proton and neutron drip lines predicted by a
recent mass formula \citep[HFB-9,][]{Gori05}, all the way from single
neutrons and protons up to the $Z = 110$ isotopes \citep[see][for more
detail]{Wana06}. The $\alpha$-decay chains and spontaneous fission
processes are taken into account only after the freezeout of all other
reactions \citep{Cowa99, Wana02}. Neutron-induced and $\beta$-delayed
fissions, as well as the contribution of fission fragments to the
lighter nuclei, are neglected.

Each nucleosynthesis calculation is initiated when the temperature
decreases to $T_9 = 9$, at which only free nucleons exist. The initial
compositions are then given by the initial electron fraction $Y_{e}$
(number of proton per nucleon). In order to mimic the hydrodynamic
results \citep{Woos94, Arco07}, $Y_{e}$ is assumed to be $Y_{e}
(L_\nu) = (Y_{e0} - Y_{e1}) (L_\nu/L_{\nu 0}) + Y_{e1}$. $Y_{e0}$ and
$Y_{e1}$ are taken to be 0.50 and 0.15, respectively. The latter is
rather smaller than that in hydrodynamic results \citep[e.g., $\sim
0.35$ in][]{Woos94}. This extreme assumption is to obtain the heavy
$r$-nuclei abundances with a typical neutron star mass ($1.4\,
M_\odot$) for the current purpose.

\section{Hot \lowercase{$r$}-Process vs. Cold \lowercase{$r$}-Process}

The nucleosynthetic yields in each $T_\mathrm{9f}$ case are
mass-averaged over the 39 wind trajectories weighted by $\dot M
(L_\nu) \mathit{\Delta} t_\mathrm{pb}$. Figure~1 compares the
mass-averaged yields with the solar $r$-process abundances as
functions of atomic mass number. We find that, for $T_\mathrm{9f} =
1.3$ (Fig.~1a), the third-peak abundances shift to the high side from
$A = 195$. In addition, the abundance curve is unacceptably jagged
compared to the solar $r$-process pattern. For $T_\mathrm{9f} = 1.0$
(Fig.~1b), the third-peak abundances fit the solar $r$-process pattern
quite well. For $T_\mathrm{9f} = 0.7$ (Fig.~1c), however, the
third-peak abundances slightly shift to the low side from $A = 195$.

This can be understood within the classical picture of $r$-process, in
which the $(n, \gamma)$-$(\gamma, n)$ equilibrium is justified. That
is, for lower $T_\mathrm{9f}$, the $r$-process path locates at the
more neutron-rich side on the nuclide chart. As a result, the
freezeout at $N=126$ takes place at lower $Z$ (i.e., lower $A$). Among
the above three cases, $T_\mathrm{9f} = 1.0$ can be regarded as a
reasonable choice to reproduce the heavy $r$-process nuclei in the
solar system. We find, however, that this interpretation does not hold
in the lower $T_\mathrm{9f}$ cases. The $T_\mathrm{9f} = 0.4$
(Fig.~1d) case reproduces the third-peak abundances even better than
the $T_\mathrm{9f} = 0.7$ case. Moreover, for $T_\mathrm{9f} = 0.1$
(Fig.~1e), the third-peak abundances nicely fit the solar $r$-process
curve. This is a consequence of the cold $r$-processing as discussed
below.

Figure~2 shows some key physical quantities as functions of time ($t =
0$ at $T_9 = 9$) in the $L_\nu = 2 \times 10^{51}\, \mathrm{erg\
s}^{-1}$ case. The asymptotic entropy per nucleon in this wind is only
92 (in units of the Boltzmann constant). However, the (artificial) low
initial $Y_e$ ($=0.22$) leads to a high neutron-to-seed ratio
$Y_n/Y_h$ ($= 109$) at the beginning of the $r$-process ($T_9 =
2.5$). Here, $Y_n$ and $Y_h$ are the abundances of free neutrons and
the heavy nuclei ($Z > 2$), respectively. At $T_9 = 2.5$, the
abundance-averaged mass number of the heavy nuclei is 103. This is
thus taken to be representative of the third-peak forming winds. On
each line in Figure~2, open and filled triangles denote the times when
$Y_n/Y_h$ decreases to 80 and 10, respectively. The former and latter
approximately correspond to the epochs at which the second ($A = 130$)
and third ($A = 195$) peaks form, respectively. Open circles denote
the time when $Y_n/Y_h$ decreases to unity, which is referred to as
``$n$-exhaustion''. No heavier nuclei are synthesized beyond this
stage, but the rearrangement of the local abundance distribution
continues.

The ratios $\tau_{n\gamma}/\tau_{\gamma n}$ and
$\tau_{\beta}/\tau_{n\gamma}$ in each case are shown in Figures~2c and
2d, where $\tau_{n\gamma}$, $\tau_{\gamma n}$, and $\tau_{\beta}$ are
the abundance-averaged mean lifetimes of $(n, \gamma)$, $(\gamma, n)$,
and $\beta$-decay for $Z > 2$ nuclei, respectively \citep[see eqs.~(3)
and (4) in][]{Wana04}. These represent the lifetimes of the dominant
species at a given time. The condition $\tau_\beta/\tau_{n\gamma} = 1$
is referred to as ``freezeout'' (\textit{filled circles}; Fig.~2). The
abundance curve fixes at this time. Note that the classical
$r$-process is characterized by the conditions
$\tau_{n\gamma}/\tau_{\gamma n} \approx 1$ and $\tau_\beta/\tau_n \gg
1$.

As can be seen in Figure~2c, the second peak forms at $t = 0.08-0.1$~s
in all cases, where $\tau_{n\gamma}/\tau_{\gamma n} \approx 1$. This
means that the second peak forms under the condition of the $(n,
\gamma)$-$(\gamma, n)$ equilibrium. For $T_\mathrm{9f} = 1.3$, this
condition holds during the whole stage of the $r$-process. For
$T_\mathrm{9f} = 1.0$, $\tau_{n\gamma}/\tau_{\gamma n}$ slightly
decreases after $n$-exhaustion and a quasi $(n, \gamma)$-$(\gamma, n)$
equilibrium continues until freezeout. For $T_\mathrm{9f} = 0.7$,
$\tau_{n\gamma}/\tau_{\gamma n}$ slightly decreases after the
second-peak formation and a quasi equilibrium continues by the
third-peak formation. In these cases, therefore, the temperature is
enough high to maintain the (quasi) $(n, \gamma)$-$(\gamma, n)$
equilibrium during the major $r$-process phase. Hereafter, the
$T_\mathrm{9f} = 1.0$ case is taken to be representative of the
$r$-process in a high temperature environment (hereafter ``hot
$r$-process'').

In contrast, in lower $T_\mathrm{9f}$ cases,
$\tau_{n\gamma}/\tau_{\gamma n}$ quickly drops below unity and the
$(n, \gamma)$-$(\gamma, n)$ equilibrium is never achieved after the
second-peak formation. This is due to the quickly decreasing $T_9$ and
$N_n$, in which the $N_n$-$T_9$ condition abruptly falls off the
``waiting point validity boundary'' \citep{Came83, Math90,
Gori96}. The temperature does not play any roles in such an
environment, which is thus designated as the cold
$r$-process. Hereafter, the $T_\mathrm{9f} = 0.1$ case is taken to be
representative of the cold $r$-process. One may consider that, without
photodisintegration, the $r$-process path approaches the neutron-drip
line and the third peak shifts to the rather low side from $A = 195$.

The reason for this misunderstanding can be found in Figure~2d, where
the lifetimes of $\beta$-decay and of neutron capture are
compared. The hot $r$-process satisfies the classical $r$-process
condition $\tau_\beta/\tau_n \gg 1$. In the cold $r$-process, however,
$\tau_\beta$ is only a few times larger than $\tau_n$, in which the
neutron capture competes with the $\beta$-decay. This is due to the
quickly decreasing $N_n$ in the supersonically expanding
wind. Consequently, the nucleosynthetic path of the cold $r$-process
is pushed back and locates at a similar position to that of the hot
$r$-process.

This is clearly seen in Figure~2e that shows the $r$-process path in
each case in terms of the neutron separation energy. Here, $S_{2n}/2$
(two-neutron separation energy divided by two) is abundance-averaged
for the $Z > 2$ nuclei ($S_\mathrm{a}$), which approximately
represents the nucleosynthetic path at a given time\footnote{The
animations for $T_\mathrm{9f}= 1.0$ and 0.1 are available from
http//supernova.astron.s.u-tokyo.ac.jp/\~{}wanajo/research.html.}. That
is, $S_\mathrm{a} = 0$ is the neutron-drip line, while a higher
$S_\mathrm{a}$ locates at closer to the $\beta$-stability. The path
obtained from the $(n, \gamma)$-$(\gamma, n)$ equilibrium condition
\citep[$S_\mathrm{a}^0$, eq.~(3) in][]{Gori96} for each is also
plotted, which poorly predicts the \textit{real} path for the cold
$r$-process. The cold $r$-process takes a significantly higher
$S_\mathrm{a}$ than $S_\mathrm{a}^0$ owing to a push-back by
$\beta$-decay. As a result, both the cold and hot $r$-processes trace
similar $S_\mathrm{a}$ histories. Notably, the freezeout in the
$T_\mathrm{9f} = 1.0$ and 0.1 cases takes place at similar
$S_\mathrm{a}$ values. As a consequence, similar $r$-process curves
appear as can be seen in Figures~1b and 1e. Note that, for
$T_\mathrm{9f} = 0.1$, freezeout occurs at $T_9 \approx 0.2$. Hence,
any $T_\mathrm{9f} < 0.2$ cases would result in the identical
abundance curve.

\section{Abundance Feature in the Cold \lowercase{$r$}-Process}

The abundance differences between the hot and cold $r$-processes
become further small when plotted as functions of atomic
number. Figure~3 compares the mass-averaged abundances with the solar
$r$-process curve and the measured elements in the $r$-process-rich
Galactic halo star CS~31082-001. All cases, except for $T_\mathrm{9f}
= 1.3$, are in reasonable agreement with the solar $r$-process and
measured patterns. This may support the robustness of the abundance
curves found in the $r$-process-rich stars. Note that the total mass
of the $r$-process nuclei ($Z \ge 90$) is the same for all cases
($\approx 2.9 \times 10^{-4}\, M_\odot$), since the seed abundances
($A \sim 100$) are already produced at $T_9 \sim 2.5$ ($>
T_\mathrm{9f}$). The mass fractions of some measurable elements (and
$^{195}$Pt) are presented in Table~1, which also shows a small
variation between the hot and cold $r$-processes (up to a factor of
two, except for $T_\mathrm{9f} = 1.3$). These differences are small
enough, when considering the uncertainties in the nuclear data as well
as in the outflow conditions.

We can see, however, a distinctive feature in the abundance curves
around $A = 200-230$ (before $\alpha$-decay; \textit{black lines} in
Fig.~1) between the hot and cold $r$-processes. This is due to their
different freezeout histories (Fig.~2e). The hot $r$-process
(\textit{red line}) follows a low $S_\mathrm{a}$ ($\approx 2$~MeV)
path at $n$-exhaustion. At this $S_\mathrm{a}$ path, a trough at $A
\approx 200$ appears in the abundance curve (before $\alpha$-decay,
Fig.~1b). This is a consequence that the nuclei at $N \approx 130-140$
are predicted to be unstable against neutron capture \citep[see the
iso-$S_{2n}/2$ curves in Fig.~6,][]{Wana04}. On the other hand, the
cold $r$-process (\textit{blue line}) takes a high $S_\mathrm{a}$
($\approx 3$~MeV) path at $n$-exhaustion (Fig.~2e). Therefore, the
trough shifts to $A \approx 210$. The trough is further eroded by
neutron capture without push-back by photodisintegration. As a
consequence, the parent nuclei of Pb ($A = 210-231$ and 234; Fig.~1e)
result in being significantly deficient. This may lead to a visible
difference in the Pb production (Fig.~3 and Table~1), despite the
similar Th+U values\footnote{There is a local problem in the odd-even
behaviors for $A > 240$ (Fig.~1) when the current nuclear data set is
utilized. This leads to an unusual pattern among $^{232}$Th,
$^{235}$U, and $^{238}$U, which however does not affect the initial Pb
abundance. Therefore, only the sum of Th and U are presented.}.

\begin{deluxetable}{ccccccccc}
\tablecaption{Mass Fraction (in units of $10^{-2}$)}
\tablewidth{0pt}
\tablehead{
\colhead{$T_\mathrm{9f}$} &
\colhead{Zr} &
\colhead{Ba} &
\colhead{Eu} &
\colhead{Os} &
\colhead{Ir} &
\colhead{$^{195}$Pt} &
\colhead{Pb} &
\colhead{Th+U}
}
\startdata
1.3 & 7.58 & 0.56 & 0.14 & 0.19 & 2.59 & 1.86 & 1.67 & 1.40 \\
1.0 & 7.97 & 1.89 & 0.20 & 2.04 & 3.24 & 1.69 & 2.06 & 1.27 \\
0.7 & 8.04 & 1.49 & 0.31 & 2.90 & 3.51 & 1.05 & 1.45 & 1.26 \\
0.4 & 8.02 & 1.32 & 0.32 & 2.71 & 3.49 & 1.25 & 0.97 & 1.31 \\
0.1 & 7.97 & 1.32 & 0.36 & 2.34 & 2.97 & 1.58 & 0.92 & 1.25
\enddata
\end{deluxetable}

\section{Implications}

In this \textit{Letter}, the $r$-process in a low temperature
environment ($T_\mathrm{9f} \sim 0.1$; cold $r$-process) was
investigated, which is characterized by the conditions
$\tau_{n\gamma}/\tau_{\gamma n} \ll 1$ and
$\tau_{\beta}/\tau_{n\gamma} \gtrsim 1$. Despite the fundamental
difference of the cold $r$-process compared to the traditional (hot)
$r$-process, a solar-like $r$-process curve can be reproduced quite
well. These physical conditions are realized in the supersonically
expanding neutrino-driven outflows. In particular, the winds from a
low-mass progenitor that has a steep core density gradient easily
expand to a large distance (and thus to low $T_\mathrm{9f}$) before
colliding to the preceding supernova ejecta \citep{Arco07}. Therefore,
the low-mass end of supernova progenitors (e.g., $8-12\, M_\odot$) may
be the most likely site of the cold $r$-process. The low-mass
supernovae have been also suggested to be the $r$-process site from
Galactic chemical evolution studies \citep{Math90, Ishi99}.

Among the six $r$-process-rich Galactic halo stars currently reported,
CS~31082-001 \citep{Plez04} and HE~1523-0901 \citep[upper
limit,][]{Freb07} appear to have the low Pb abundances compared to the
scaled solar $r$-process curve. The upper limit of Pb for
BD~+17$^\circ$3248 \citep{Cowa02} is still consistent to the scaled
solar value. The other three, CS~22892-052 \citep{Sned03}, HD~115444
\citep{West00}, and HD~221170 \citep{Ivan06} have the measured Pb
abundances consistent to the (high) scaled solar value. The variation
of the Pb production may be attributed to the different types of (cold
and hot) $r$-processes. The low Pb abundance in CS~31082-001 has been
in fact a most worrisome aspect, when the U/Th value is applied for
the age dating \citep{Plez04}. This is hardly explained from the
classical (or hot) $r$-process view, in which a low $S_\mathrm{a}$
($\approx 2$~MeV) event is inevitable for the third-peak formation. In
this regard, the cold $r$-process might enables us to consistently
treat the Pb, Th, and U abundances for the cosmochronology.

Finally, it is noted that the cold $r$-process brings us an additional
challenge to the theoretical $r$-process study. In most of previous
$r$-process calculations, the uncertainties in neutron-capture rates
have not been considered seriously. This is reasonable for the hot
$r$-process, in which uncertainties even by a factor of ten would not
cause a serious problem (see Fig.~2d). In the cold $r$-process,
however, the neutron capture competes with the $\beta$-decay, and thus
accurate neutron-capture (as well as $\beta$-decay) rates will be
required to predict reliable $r$-process abundances.

\acknowledgements

I would like to acknowledge an anonymous referee for helpful
suggestions. This work was supported in part by a Grant-in-Aid for
Scientific Research (17740108) from the Ministry of Education,
Culture, Sports, Science, and Technology of Japan.

\end{document}